\documentclass[11pt,a4paper]{article}

\usepackage{graphicx}

\usepackage{amsmath}
\usepackage{verbatim}
\usepackage{amssymb}
\usepackage{hyperref}
\usepackage{comment}
\usepackage{enumitem}
\newlist{contract}{enumerate}{10}
\setlist[contract]{label*=\arabic*.}
\setlistdepth{10} 

\usepackage{amsthm}
\usepackage{geometry}
\usepackage{float}

\usepackage{setspace}
\spacing{1.5}
\usepackage{fancyhdr}

\usepackage{color} 

\usepackage{mathabx}
\usepackage{overpic}

\usepackage[all]{xy}

\graphicspath{{pics/}}

\newcommand{\bpf}{\begin{proof}}

\newcommand{\epf}{\end{proof}}

\newcommand{\bthm}{\begin{thm}}

\newcommand{\ethm}{\end{thm}}

\newcommand{\bprop}{\begin{prop}}

\newcommand{\eprop}{\end{prop}}

\newcommand{\bcor}{\begin{cor}}

\newcommand{\ecor}{\end{cor}}

\newcommand{\blem}{\begin{lem}}

\newcommand{\elem}{\end{lem}}

\newcommand{\bdefn}{\begin{defn}}

\newcommand{\edefn}{\end{defn}}

\newcommand{\bexmp}{\begin{exmp}}

\newcommand{\eexmp}{\end{exmp}}

\newcommand{\brem}{\begin{rem}}

\newcommand{\erem}{\end{rem}}

\newcommand{\bdia}{\begin{displaymath}\xymatrix}

\newcommand{\edia}{\end{displaymath}}

\newcommand{\beq}{\begin{equation*}\begin{aligned}}

\newcommand{\eeq}{\end{aligned}\end{equation*}}

\newtheorem{thm}{\textbf {Theorem}}[section]

\newtheorem{cor}[thm]{\textbf{Corollary}}

\newtheorem{prop}[thm]{\textbf{Proposition}}

\newtheorem{lem}[thm]{\textbf{Lemma}}

\theoremstyle{definition}

\newtheorem{defn}[thm]{\textbf{Definition}}

\newtheorem{exmp}[thm]{Example}

\newtheorem{notn}[thm]{Notation}

\newtheorem{conv}[thm]{Convention}

\theoremstyle{remark}

\newtheorem{rem}[thm]{Remark}

\usepackage{hyperref}
\usepackage{float}

\DeclareMathOperator{\vcong}{vcong}
\DeclareMathOperator{\econg}{econg}
\DeclareMathOperator{\cg}{cong}
\DeclareMathOperator{\length}{length}
\DeclareMathOperator{\vol}{vol}
\DeclareMathOperator{\size}{size}
\DeclareMathOperator*{\argm}{argmin}

\author{A thesis presented by \\ Nithin Kavi \\ \\ to \\ Computer Science \\ in partial fulfillment of the joint concentration \\requirements for the degree of Bachelor of Arts \\ in Computer Science and Mathematics \\ \\ Harvard College \\ Cambridge, Massachusetts \\ April 1, 2024}

\title{Partial Implementation of Max Flow and Min Cost Flow in Almost-Linear Time}
\date{}
\begin{document}
\bibliographystyle{plain}
\maketitle
\thispagestyle{empty}

\newpage
\vfill
\begin{abstract}
\thispagestyle{empty}

In 2022, Chen et al. proposed an algorithm in \cite{main} that solves the min cost flow problem in $m^{1 + o(1)} \log U \log C$ time, where $m$ is the number of edges in the graph, $U$ is an upper bound on capacities and $C$ is an upper bound on costs. However, as far as the authors of \cite{main} know, no one has implemented their algorithm to date. In this paper, we discuss implementations of several key portions of the algorithm given in \cite{main}, including the justifications for specific implementation choices. For the portions of the algorithm that we do not implement, we provide stubs. We then go through the entire algorithm and calculate the $m^{o(1)}$ term more precisely. Finally, we conclude with potential directions for future work in this area.
 
\end{abstract}

\newpage
\thispagestyle{empty}
\section*{Acknowledgments}

I would like to thank Dr. Adam Hesterberg for offering guidance throughout this project as my senior thesis advisor during the 2023-2024 academic year. He was extremely helpful in discussing ideas and offering feedback on this paper, and I could not have done this project without him. I would also like to thank Professors Yang Liu, Sushant Sachdeva and Richard Peng for answering clarifying questions about the algorithm that they present in \cite{main} and offering advice on how I go about implementing things in this project. Their insight was extremely helpful. I am also thankful for Professors Daniel Sleator and Robert Tarjan for their insight on Dynamic Trees. I also sincerely appreciate Professor Madhu Sudan for agreeing to read and evaluate my thesis. 

In addition, I want to thank Mincheol, Kevin, Ryan, Skyler, Sunera, Wittmann, and Yiting for being my suitemates and being part of an unforgettable experience at Harvard. I've really enjoyed these past 4 years with them. I am also thankful to my parents and older sister for all of their support throughout my Harvard education.

Lastly, I thank the Harvard Computer Science Department and Harvard College for giving me the opportunity to write a senior thesis and undertake this project.

\newpage

\tableofcontents
\thispagestyle{empty}
\section{Introduction}

The maximum-flow problem is formulated as follows: 

\textit{Suppose we have a connected graph $G$ with $m$ edges and $n$ vertices where one vertex is designed as a source $s$ and another is designated as a sink $t$. Each edge $e$ in $G$ has a given capacity $c_e.$ We want to the maximize the amount of flow from $s$ to $t$ such that for all vertices $v \neq s, t$ the flow out of $v$ is equal to the flow into $v$ and that for any edge $e$, the flow $f_e$ along edge $e$ satisfies $f_e \leq c_e.$} 

To solve this problem, Ford and Fulkerson created the Ford-Fulkerson algorithm, which runs in $O(mf^*)$ time where $f^*$ is the max flow. There is also the Edmonds-Karp algorithm which runs in $O(nm^2)$ time which does not depend on the size of the max flow itself. 

There is also the related minimum-cost flow problem, where each edge is also assigned a cost along with its capacity. Then the goal is to push the maximum total flow from the source to the sink while minimizing the total cost along the way.

Lastly, we can combine these two ideas as follows: Letting the max flow as defined above be $f^*,$ the min cost flow finds the minimum possible total cost of the flow over the edges such that all requirements of the max flow problem are met and the max flow is attained.

In 2022, Chen, Kyng, Liu, Peng, Gutenberg and Sachdeva wrote \cite{main} in which they found an algorithm to compute both maximum flows and minimum-cost flows in $m^{1 + o(1)} \log U \log C$ time, which is the asymptotically fastest known algorithm to date. In this paper, we work on implementing the algorithm given in \cite{main}. As far as the authors of \cite{main} are aware, this algorithm has never been implemented before, and this paper constitutes the first attempt at implementing it. This algorithm relies on a number of subparts, including dynamic trees, low-stretch spanning trees, and undirected minimum-ratio cycles. Apart from very well-known things (like binary trees), there do not appear to be existing implementations of any of these subparts that meet the asymptotic time complexities given in \cite{main}.

We implement some of these subparts in the following sections in Python. Although Python is slower than many other languages by a large constant factor, the algorithm in \cite{main} was never designed to be fast in practice, so this is not an issue. At the beginning of each section, we briefly describe what code that section corresponds to and explain how that code fits into the broader algorithm. In addition, for the sections where we do not implement all of the code described in the corresponding subpart, we include a description of what we did implement and what was challenging about the parts we did not implement. All of the code referenced in this paper can be found at the following Github: \url{https://github.com/thinkinavi24/almost_linear_max_flow}. The generative AI tool ChatGPT was used to write a significant amount of this code and is credited accordingly in comments of the Jupyter codeblocks. In multiple Jupyter notebooks, some codeblocks were written by ChatGPT while others were written by hand; any code block not labeled as being written by ChatGPT was written by hand. All code written by generative AI was thoroughly checked, tested and revised as needed as was all other code in the Github.

\section{Preliminaries}

We begin with some definitions that we will refer back to throughout this paper. Definitions, notations and remarks with references in this section are copied identically from those references and are reproduced here for convenience. In addition, in later sections throughout this paper, we will quote results including theorems, definitions and lemmas from other articles and we will mention the source of the result immediately before reproducing it. Further, when discussing the implementation of a list of methods/attributes in a data structure that are described in another paper, we will often quote that paper's description of the methods/attributes. When we do so, we will mention the source of the result immediately before reproducing a list that is copied from the referenced paper. 

\subsection{Definitions}

\begin{defn}
    Given a graph $G,$ for all pairs of vertices $(u, v) \in G$ we define $d_G(u, v)$ to be the shortest path distance from $u$ to $v$ on $G.$
\end{defn}

\begin{defn} (\cite{AN19}) The \textit{radius} of a connected graph $G$ around a point $x_0 \in G$ is $\max_{v \in G} d(x_0, v)$ where $d(x_0, v)$ represents the shortest distance from $x_0$ to $v$ for each vertex $v \in G.$ 
\end{defn}

\begin{notn}\label{B_defn} (\cite{main}) We let a graph $G$ correspond to its vertices $V$ and its edges $E$, and we let $n = |V|$  and $m = |E|.$ Sometimes, we may explicitly write $E(G)$ to denote the edges of $G$ or $V(G)$ to denote the vertices of $G.$ Further, we let $e^G \in E(G)$ denote an edge of $G$ and $v^G \in V(G)$ denote a vertex of $G.$ We assume that each edge $e \in E$ has an implicit direction, used to define its edge-vertex incidence matrix $\mathbf{B}.$ 
    
\end{notn}

\begin{defn} (\cite{main}) We let $\deg_G(v)$ denote the degree of the vertex $v$ in $G,$ or the number of incident edges. We let $\Delta_{\max}(G)$ and $\Delta_{\min}(G)$ denote the maximum and minimum degree of graph $G.$
    
\end{defn}

\begin{notn} For a graph $G$ with vertices $V,$ for any subset $S \subset V,$ we let $\overline S$ denote $V \backslash S.$
    
\end{notn}

\begin{defn} (\cite{main}) We define the volume of a set $S \subseteq V$ as $\vol_G(S) = \sum_{v \in S} \deg_G(v).$
    
\end{defn}

\begin{defn} (\cite{main}) Given graphs $G$ and $H$ with $V(G) \subseteq V(H)$ we say that $\Pi_{G \to H}$ is a \textit{graph-embedding} from $G$ into $H$ if it maps each edge $e^G = (u, v) \in E(G)$ to a $u-v$ path $\Pi_{G \to H}(e^G)$ in $H$. 
    
\end{defn}

\begin{defn}\label{econg} (\cite{main}) The congestion of an edge $e^H$ is defined as $\econg(\Pi_{G \to H}, e^H) = |\{e^G \in E(G) | e^H \in \Pi_{G \to H}(e^G) \}|$ and of the embedding by $\econg(\Pi_{G \to H}) = \max_{e^H \in E(H)} \econg(\Pi_{G \to H}, e^H).$
    
\end{defn} 

\begin{defn}\label{vcong} (\cite{main}) The congestion of a vertex $v^H \in V(H)$ is defined by $\vcong(\Pi_{G \to H}, v^H) = |\{e^G \in E(G) | v^H \in \Pi_{G \to H}(e^G) \}|$ Further, the vertex-congestion of the embedding is $\vcong(\Pi_{G \to H}) = \max_{v^H \in V(H)} \vcong(\Pi_{G \to H}, v^H).$
    
\end{defn}

\begin{defn}\label{length} (\cite{main}) The length of an embedding is defined as $\length(\Pi_{G \to H}) = \max_{e^G \in E(G)} |\Pi_{G \to H}(e^G)|.$
    
\end{defn}

\begin{defn} (\cite{main}) $G$ is a dynamic graph if it undergoes batches $U^{(1)}, U^{(2)}, \ldots$ of updates consisting of edge insertions/deletions and/or vertex splits that are applied to $G.$ We say that the graph $G$, after applying the first $t$ update batches $U^{(1)}, U^{(2)}, \ldots, U^{(t)}$ is at \textit{stage} $t$ and denote the graph at this stage by $G^{(t)}.$
    
\end{defn}

\begin{rem} (\cite{main}) For each update batch $U^{(t)},$ we encode edge insertions by a tuple of tail and head of the new edge and deletions by a pointer to the edge that is about to be deleted. We further also encode vertex splits by a sequence of edge insertions and deletions as follows: if a vertex $v$ is about to be split and the vertex that is split off is denoted $v^{\text{NEW}}$, we can delete all edges that are incident to $v$ but should be incident to $v^{\text{new}}$ from $v$ and then re-insert each such edge via an insertion (we allow insertions to new vertices that do not yet exist in the graph). For technical reasons, we assume that in an update batch $U^{(t)}$, the updates to implement the vertex splits are last, and that we always encode a vertex split of $v$ into $v$ and $v^{\text{NEW}}$ such that $\deg_{G^{(t + 1)}}(v^{\text{NEW}}) \leq \deg_{G^{(t + 1)}}(v).$ We let the vertex set of graph $G^{(t)}$ consist of the union of all endpoints of edges in the graph (in particular if a vertex is split, the new vertex $v^{\text{NEW}}$ is added due to having edge insertions incident to this new vertex $v^{\text{NEW}}$ in $U^{(t)}).$
    
\end{rem}

\begin{defn}\label{tilde} We say a function $g(n) = \tilde O(f(n))$ if there exist constants $c, k$ such that $g(n) \leq c f(n) \log^k(n)$ for sufficiently large $n.$ Note that this is equivalent to standard big O-notation except we also allow for arbitrary powers of $\log$.
     
\end{defn}

\begin{defn} (\cite{main}) We define ENC($u$) of an update $u \in U^{(t)}$ to be the size of the encoding of the update and note that for edge insertions/deletions, we have ENC($u$) = $\tilde O(1)$ and for a vertex split of $v$ into $v$ and $v^{\text{NEW}}$ as described above we have that ENC($u$) = $\tilde O(\deg_{G^{(t + 1)}}) (v^{\text{NEW}}).$ For a batch of updates $U$, we let ENC($U$) = $\sum_{u \in U}$ ENC($u$).
    
\end{defn}

\begin{rem} (\cite{main}) The number of updates $|U|$ in an update batch $U$ can be completely different from the actual encoding size ENC $(U)$ of the update batch $U$.
    
\end{rem}

\begin{defn} (\cite{sw19}) For any subsets $S, T \subset V$ we denote $E(S, T) = \{\{u, v\} \in E | u \in S, v \in T\}$ as the set of edges between $S$ and $T.$
    
\end{defn}

\begin{defn} (\cite{sw19}) The cut-size of a cut $S$ is $\delta(S) = |E(S, \overline S)|.$
    
\end{defn}

\begin{defn}\label{conductance} (\cite{sw19}) The \textit{conductance} of a cut $S$ in $G$ is $\Phi_G(S) = \frac{\delta(S)}{\min(\vol_G(S), \vol_G(V \backslash S))}.$ Unless otherwise noted, when speaking of the conductance of a cut $S,$ we assume $S$ to be the side of smaller volume. The conductance of a graph $G$ is $\Phi_G = \min_{S \subset V} \Phi_G(S).$ If $G$ is a singleton, we define $\Phi_G = 1.$
     
\end{defn}

\begin{defn}\label{expander} (\cite{sw19}) We say a graph $G$ is a $\phi$ \textit{expander} if $\Phi_G \geq \phi$, and we call a partition $V_1, \ldots, V_k$ of $V$ a $\phi$ expander decomposition if $\min_i \Phi_{G_{[V_i]}} \geq \phi.$
    
\end{defn}

\begin{defn}\label{near_expander} (\cite{sw19}) Given $G = (V, E)$ and a set of vertices $A \subset V,$ we say $A$ is a nearly $\phi$ expander in $G$ if $\forall S \subseteq A, \vol(S) \leq \vol(A)/2 : |E(S, V \backslash S)| \geq \phi \vol(S).$
    
\end{defn}

\begin{defn}\label{subdivision} (\cite{sw19}) Given a graph $G = (V, E)$, its subdivision graph $G_E = (v', E')$ is the graph where we put a split [vertex] $x_e$ on each edge $e \in E$ (including self loops). Formally, $V' = V \cup X_E$ where $X_e = \{x_e | e \in E \}$ and $E' = \{\{u, x_e\}, \{v, x_e\} |e = \{u, v\} \in E\}.$ We will call [vertices] in $X_E$ the split [vertices], and the other [vertices] in $V'$ the regular [vertices]. 

\begin{conv} Whenever we use $\log$ in this paper, we always mean the natural logarithm. Of course, this is irrelevant when considering only asymptotic runtimes.
    
\end{conv}
    
\end{defn}

\subsection{Layout of Algorithm}\label{layout}

The authors in \cite{main} present their main algorithm on page 69 at the beginning of Section 9 in Algorithm 7. In order to figure out exactly what needs to be implemented (vs portions of \cite{main} that are not directly part of the final algorithm) we work backwards from Algorithm 7. 

We begin by listing the things we need to implement for Algorithm 7. Then for each of these, we list the things we need to implement for them, and so on until we reach algorithms that can be implemented without referring to any other major algorithms. Note that some things are repeated within the list; this is because certain data structures/algorithms are used at multiple points within the overall main algorithm. The bolded items are ones which we have implemented and are labeled as fully implemented or partially implemented.

A quick glance at Algorithm 7 tells us that we need to implement the following:

\begin{contract}
    \item The data structure $\mathcal{D}^{HSFC}$ (Theorem 6.2, \cite{main})
    \begin{contract}
        \item Branching Tree Chain (Theorem 7.1, \cite{main})
        \begin{contract}
            \item Dynamic Low Stretch Decomposition (LSD) (Lemma 6.5, \cite{main}) (partial implementation in Section \ref{lsd_section})
            \begin{contract}
                \item \textbf{Low-stretch spanning trees} (Theorem 3.2, \cite{main}) (full implementation in Section \ref{lsd_section})
                \begin{contract}
                    \item \textbf{create\_petal algorithm} (\cite{AN19})
                    \item \textbf{petal\_decomposition algorithm} (\cite{AN19})
                    \item \textbf{hierarchical\_petal\_decomposition algorithm} (\cite{AN19})
                \end{contract}
                \item \textbf{Heavy Light Decomposition} (Lemma B.8, \cite{main}) (full implementation in Section \ref{lsd_section})
                \item Rooted tree (Lemma B.9, \cite{main})
                \item Computing Stretch Overestimates
                \begin{contract}
                    \item Computing permutation $\pi$ (Lemma B.6, \cite{main})
                    \begin{contract}
                        \item Compute congestion (Lemma B.5, \cite{main})
                        \begin{contract}
                            \item \textbf{Dynamic Trees} (Lemma 3.3, \cite{main}) (external implementation cited in Section \ref{dynamic_trees_section})
                        \end{contract}
                    \end{contract}
                \end{contract}
                \item Tree Decomposition (Lemma B.7 of \cite{main}, quotes \cite{st03} and \cite{st04})
            \end{contract}
            \item Sparsified core graph (Definition 6.9, \cite{main}), implemented in Algorithm 4 (Lemma 7.8. \cite{main})
                \begin{contract}
                    \item Dynamic Spanner (Theorem 5.1, \cite{main})
                    \begin{contract}
                        \item SPARSIFY algorithm (Theorem 5.5, \cite{main})
                        \begin{contract}
                            \item DECOMPOSE algorithm (Theorem 5.11, \cite{main})
                            \begin{contract}
                                \item Expander Decomposition (Theorem 1.2 of \cite{sw19})
                                \begin{contract}
                                    \item Cut-Matching (Theorem 2.2 of \cite{sw19})
                                    \begin{contract}
                                        \item Implicitly update flow $f$ (Appendix B.1 of \cite{sw19})
                                    \end{contract}
                                    \item Trimming (Theorem 2.1 of \cite{sw19})
                                \end{contract}
                            \end{contract}
                            \item Data structure $\mathcal{DS}_{ExpPath}$ (Theorem 5.12 of \cite{main}, quotes \cite{cs21})
                            \begin{contract}
                                \item Expander Pruning (Theorem 1.3 of \cite{sw19})
                                \item Theorem 3.8 of \cite{cs21}
                                \item ES-Tree Data Structure (\cite{es-tree})
                            \end{contract}
                        \end{contract}
                        \item Update algorithm
                    \end{contract}
                    \item Dynamic Core Graphs (Lemma 7.5, \cite{main})
                    \begin{contract}
                        \item Dynamic Low Stretch Decomposition (LSD) (Lemma 6.5, \cite{main})
                        \item Multiplicative Weights Update (MWU) (Lemma 6.6, \cite{main})
                    \end{contract}
                \end{contract} 
        \end{contract}
        \item \textbf{Rebuilding Game} (Lemma 8.3, \cite{main}) (full implementation in Section \ref{rebuilding_section})
    \end{contract}
    \item \textbf{Dynamic Trees} (Lemma 3.3, \cite{main}) (external implementation in Section \ref{dynamic_trees_section})
\end{contract}

In order to implement Algorithm 7, we can view this list itself as a flowchart/DAG, where the leaves of the tree need to be implemented to eventually implement our way up to the root. In order to implement the algorithm at each ``node" of the tree, all of its children must be implemented. 

In Section \ref{overall}, we analyze these different components and the asymptotic guarantees given in \cite{main} to precisely calculate the asymptotic time complexity of the main algorithm in \cite{main}. Lastly, we conclude the paper in Section \ref{conclusion} with a summary of the work and possible directions for future study.



\section{Dynamic Trees}\label{dynamic_trees_section}


Dynamic Trees are described in Lemma 3.3 of \cite{main} which is reproduced below for convenience:

\begin{lem}\label{dynamic_trees_lemma} (\cite{main}) \textit{There is a deterministic data structure $\mathcal{D}^{(T)}$ that maintains a dynamic tree $T \subseteq G = (V, E)$ under insertion/deletion of edges with gradients $\mathbf{g}$ and lengths $\mathbf{\ell}$ and supports the following operations:}

\begin{enumerate}
    \item \textit{Insert/delete edges $e$ to $T$, under the condition that $T$ is always a tree, or update the gradient $\mathbf{g_e}$ or lengths $\mathbf{\ell_e}.$ The amortized time is $\tilde O(1)$ per change.}

    \item \textit{For a path vector, $\Delta = \mathbf{p}(T[u, v])$ for some $u, v \in V$, return $\langle \mathbf{g}, \mathbf{\Delta} \rangle$ or $\langle \mathbf{\ell}, |\mathbf{\Delta}| \rangle$ in time $\tilde O(1).$}

    \item \textit{Maintain a flow $\mathbf{f} \in \mathbb R^E$ under operations $\mathbf{f} \leftarrow \mathbf{f} + \eta \Delta$ for $\eta \in \mathbb R$ and path vector $\Delta = \mathbf{p}(T[u, v])$ or query the value $\mathbf{f}_e$ in amortized time $\tilde O(1).$}

    \item \textit{Maintain a positive flow $\mathbf{f} \in \mathbb R^E$ under operations $\mathbf{f} \leftarrow \mathbf{f} + \eta |\Delta|$ for $\eta \in \mathbb R_{\geq 0}$ and path vector $\mathbf{\Delta} = \mathbf{p}(T[u, v])$, or query the value $\mathbf f_e$ in amortized time $\tilde O(1)$.}

    \item \textit{DETECT(). For a fixed parameter $\varepsilon$, and under positive flow updates (item 4), where $\Delta^{(t)}$ is the update vector at time $t$, returns $S^{(t)} = \left\{e \in E : \mathbf{\ell_e} \sum_{t' \in [\textbf{last}_e^{(t)} + 1, t]} |\Delta_e^{(t')}| \geq \varepsilon \right\}$ where $\text{last}_e^{(t)}$ is the last time before $t$ that $e$ was returned by DETECT(). Runs in time $\tilde O(|S^{(t)}|).$}
\end{enumerate} 

\end{lem}

The data structure in Lemma 3.3 of \cite{main} above is a slight extension on the Dynamic Trees described in \cite{st83}. In general, the purpose of dynamic trees is to make operations like cutting a tree or linking two trees possible in $\tilde O(1)$ time, which can then be efficiently combined to implement operations 1-5 above also in $\tilde O(1)$ time.

This is an important data structure for the algorithm presented in \cite{main}. It is challenging to find an implementation of this data structure online. The authors of \cite{st83} pointed us to the following implementation in C++: \url{https://codeforces.com/contest/487/submission/8902202}, written by the first author of \cite{st83}. They also pointed us to their paper \cite{st85} which provides a simpler description for Dynamic Trees.

Since an implementation of Dynamic Trees already exists, though it is in a different language, we do not implement one here and instead defer to that implementation. While that implementation does not include a DETECT() method, the authors in \cite{main} give an algorithm for DETECT() which should be implementable given the Dynamic Tree Class linked above. Even though we are not creating our own dynamic trees implementation, we will refer to Lemma \ref{dynamic_trees_lemma} throughout this paper so it is important to understand their definition as given by Lemma \ref{dynamic_trees_lemma}.

\section{Low Stretch Decomposition}\label{lsd_section}

As described in Section \ref{layout}, in this section we implement Low Stretch Decomposition as specified in Lemma 6.5 of \cite{main}, which states the following:

\begin{lem}\label{lem6.5} (Dynamic Low Stretch Decomposition). There is a deterministic algorithm with total runtime $\tilde O(m)$ that on a graph $G = (V, E)$ with lengths $\mathbf{\ell} \in \mathbb R^E_{> 0},$ weights $\mathbf v \in \mathbb R^E_{> 0},$ and parameter $k$, initializes a tree $T$ spanning $V,$ and a rooted spanning forest $F \subseteq T,$ a edge-disjoint partition $\mathcal W$ of $F$ into $O(m/k)$ sub trees and stretch overestimates $\widetilde{\text{str}_e}.$ The algorithm maintains $F$ decrementally against $\tau$ batches of updates to $G,$ say $U^{(1)}, U^{(2)}, \ldots, U^{(\tau)},$ such that $\widetilde{\text{str}_e} ~~ \stackrel{\text{def}}{=} 1$ for any new edge $e$ added by either edge insertions or vertex splits, and:

\begin{enumerate}
    \item $F$ has initially $O(m/k)$ connected components and $O(q \log^2 n)$ more after $t$ update batches of total encoding size $q ~ \stackrel{def}{=} \sum_{t = 1}^\tau ENC(U^{(i)})$ satisfying $q \leq \tilde O(m).$

    \item $\text{str}_e^{F, \ell} \leq \widetilde{\text{str}}_e \leq O(k \gamma_{LSST} \log^4 n)$ for all $e \in E$ at all times, including inserted edges $e.$

    \item $\sum_{e \in E^{(0)}} v_e \widetilde{\text{str}}_e \leq O(||\mathbf v||_1 \gamma_{LSST} \log^2 n)$

    \item Initially, $\mathcal{W}$ contains $O(m/k)$ subtrees. For any piece $W \in \mathcal{W}, W \subseteq V, |\partial W| \leq 1$ and $\vol_G(W \backslash R) \leq O(k \log^2 n)$ at all times, where $R \supseteq \partial \mathcal{W}$ is the set of roots in $F.$ Here, $\partial W$ denotes the set of boundary vertices that are in multiple partition pieces. 
\end{enumerate}
    
\end{lem}

The factor $\gamma_{LSST}$ will be specified in Notation \ref{lsst_notn}.

We begin by implementing low-stretch spanning trees in Section \ref{lsst_section} before moving onto the Heavy Light Decomposition in Section \ref{heavy_light_section}. 

\subsection{Low-stretch spanning trees}\label{lsst_section}

We begin by writing code to implement the LSST algorithm described in \cite{AN19}. The code can be found in the file titled petal.ipynb.

The main result of \cite{AN19} can be summarized as follows:

\begin{thm}\label{main_lsst} Suppose $G$ is a connected, undirected graph with vertices $V$ and edges $E$ where $|V| = n$ and $|E| = m$. Then we can construct a spanning tree $T$ such that $\sum_{(u, v) \in E} \frac{d_T(u, v)}{w(u, v)} = O(m \log n \log \log n).$ Further, this tree can be found in $O(m \log n \log \log n)$ time. 
\end{thm}

The tree $T$ found in Theorem \ref{main_lsst} is our low-stretch spanning tree (LSST).

\begin{notn}\label{lsst_notn}
    Following the notation of \cite{main}, we will be referring to the stretch factor $O(m \log n \log \log n)$ of the LSST as $\gamma_{LSST}$ for the remainder of the paper. 
\end{notn}

The algorithm can be understood through a few functions: create\_petal, petal\_decomposition and hierarchical\_petal\_decomposition. We describe the implementation of each of these functions below. In order to get our LSST, we call hierarchical\_petal\_decomposition on our original graph with specific parameters that are specified in Section 3.3.

\subsubsection{create\_petal}

Abraham and Neiman give two algorithms in \cite{AN19} for create-petal which both take in a graph, a starting point $x_0,$ a target $t$ and a radius $r$ (where $r$ is expressed as a range in the first algorithm) and which both return a cluster (which is the petal) as well as a center of that petal. The algorithm given on page 235 is not fast enough for the overall desired bound in this paper: it requires that the ratio between the largest and smallest distances is $O(n^k)$ for some $k$ which does not necessarily hold in general, and the create-petal algorithm has a for loop which could potentially run for many iterations, and there is no clear bound on how many iterations this would take. For this reason, we implement the faster algorithm on page 245 in section 6 of \cite{AN19}, which the authors themselves state is necessary for the best asymptotic efficiency.


In the code, we begin with an ordinary implementation of Dijkstra's Algorithm using Fibonacci Heaps along with a corresponding shortest\_path function. Everything in this portion is standard. After that, we write a function to generate the directed graph $\Tilde{G}$ as specified on page 245 of \cite{AN19}. As Abraham and Neiman point out, all edges on the directed graph have nonnegative length. 

With that out of the way, we can write the create\_petal function. In essence, this algorithm iteratively carves petals off of the graph and then carves petals off of the remaining part of the graph until all petals have been carved, when any remaining vertices are made part of the stigma of the flower. This function takes in the following inputs: a graph $Y$, a target $t$, a starting point $x_0$ and a radius $r.$ Note that this is slightly different from the slower function that Abraham and Neiman propose. In particular, the create\_petal function does not take in the original graph as input except at the beginning; rather, as described in the following subsection, each time a petal is carved, create\_petal is iteratively called on the remaining subgraph. We also do not have any $\text{hi}$ and $\text{lo}$ parameters, as the faster algorithm described on page 245 of \cite{AN19} does not include any.

To create a petal, we first run Dijkstra on the given graph with the given starting point $x_0.$ We then generate the corresponding directed graph based on the shortest distances from $x_0$ to all other points in the graph and calculate the shortest path to the target $t.$ Next, we need to find the special vertex $r'$, defined on page 245 of \cite{AN19} as the furthest point from $t$ on the shortest path from $t$ to $x_0$ in the original graph $G$ that is within distance $r$ of $t.$ To do this, we could perform binary search on the sorted distances on the shortest path from $t$ to $x_0$ in order to find the largest number less than or equal to $r.$ However, in our implementation we simply do a linear search as this does not impact the overall asymptotic time complexity, which we empirically check later.

Next, the algorithm stipulates that the petal is the ball of radius $r/2$ around $t$ in the directed graph where each edge on the path from $t$ to $p_{r'}$ is set to half of its original weight in $G.$ At first, this may seem to be equivalent to the following: include every vertex within the ball of radius $r/2$ and then also include every vertex on the path from $t$ to $p_{r'}$ including $p_{r'}.$ This would make intuitive sense because $p_{r'}$ is defined as the furthest point along the path from $t$ to $x_0$ within distance $r$ of $t$. If all of the closest points to $t$ are further than $r$ away, then $p_{r'} = t.$ An implementation which does this would not be quite right for the following reason: while edges in the directed graph all have nonnegative length by the triangle inequality, they can have length 0, and this fact influences the implementation of create\_petal. Therefore, in the directed graph, $p_{r'}$ may not be the furthest point within distance $r.$ Edges of length 0 mean that some points ``further" along the path may be of equal distance.  

Therefore, it is essential to first set the weight of each edge on the path from $t$ to $p_{r'}$ to half its original weight in $G.$ After that, we run Dijkstra on the directed graph $\Tilde{G}$ and select all vertices within distance $r/2$ of $t.$

The function then returns the petal and the point $p_{r'}$ which is the center of the petal and plays a key role in the petal decomposition. In addition, we also choose to return the shortest path from $x$ to $t$, since this will come up in the petal\_decomposition function described in the next subsection. 

\subsubsection{petal\_decomposition}

In our implementation of this function, we begin with some sanity checks that the center $x$ and target $t$ are both in the graph. While this may seem unnecessary, we will see in the next subsection that there are many recursive calls which involve modifying the graph. These assert and raise statements helped catch many errors earlier on in the implementation and ensured everything worked properly once those errors were corrected. 

We begin by computing the radius of graph, defined as the maximum length across all vertices $v$ of the shortest path from $x_0$ to $v$ found using Dijkstra. We then initialize the following lists with their defined purposes:

\begin{itemize}
    \item Ys: $Y_j$ is defined as the original graph minus the vertices and edges contained in petals $X_1, \ldots, X_j$. In particular, $Y_0$ is initialized to the entire graph before any petals have been removed.
    \item Xs: $X_j$ is the petal found on iteration $j$ and is carved from $Y_{j-1}$, except for $X_0$ which is the remaining part of the graph after all petals have been carved (what the authors in \cite{AN19} call the ``stigma" of the flower)
    \item xs: $x_j$ is the center of petal $X_j,$ or the point $p_{r'}$ when computing the petal $X_j$, except for $x_0$ which is the original input 
    \item ts: $t_j$ is the target when creating petal $X_j$
    \item ys: $y_j$ is the neigbhor of $x_j$ on the path from $x_0$ to $t_j$ which is closer to $x_0$
\end{itemize} 

The rest of the implementation follows the pseudocode. The first petal that is carved may be a special petal if the distance from $x_0$ to the target $t$ is greater than $\frac{5}{8}$ of the radius from $x_0.$ Abraham and Neiman describe why they choose the $\frac{5}{8}$ constant in \cite{AN19}. The purpose of this petal is to preserve the shortest path from $x_0$ to $t.$

To carve all of the other petals, we iteratively check for points whose distance from $x_0$ is at least $\frac{3r}{4}.$ If no other points exist, we have found all petals. Otherwise, we arbitrarily choose one of those points to be the target $t_j$ for petal $X_j.$ In our implementation, we simply choose the first key in the dictionary of key value pairs where the keys are vertices and the values are distances from the center $x_0$, but this is arbitrary because the keys are only ordered as they were in the original input graph, which is arbitrary.

We then call create\_petal on the remaining part of the original graph (denoted by the element most recently appended to the Ys list) along with our center $x_0$ and target $t_j.$ For the radius, we use $r/8$. This choice comes from the pseudocode for petal\_decomposition in \cite{AN19}, where the call to create\_petal sets lo = 0 and hi = $r/8.$ However, as we recall from the previous subsection, the faster implementation of create\_petal does not have any lo or hi parameters but rather just a single radius parameter. For this reason, it was not obvious whether to use $r/8$ or something like $r/16.$ However, we decided to use $r/8$ because this parameter is a radius and all points within this radius will have distance less than or equal to $r/8$ from the center. For this reason, having a max possible value of $r/8$ (the hi parameter) vs a single value of $r/8$ seemed more analogous. However, that is not a fully rigorous argument, and the testing described in Section 4.1.4 is what provides empirical validation for this choice. 

Once we create the petal, we subtract it off from the remaining part of the graph. Using the path from $x_0$ to $t_j$, we compute $y_j$, which will be used in the following subsection. We then halve all of the edges on the shortest path to $t$ (but beginning at $x_j$ rather than $x_0$) returned by the create\_petal function as given in the pseudocode. 

Finally, we let $X_0$ be the last $Y_j$, namely the remaining part of the graph after all petals have been carved. 

\subsubsection{hierarchical-petal-decomposition}

This is the third and final function necessary to generate a low-stretch spanning tree. For inputs, it takes in the graph, a starting point $x_0$ and a target $t.$ This function is recursive with a base case that the tree on a graph with one vertex is just the single vertex itself. 

On each recursive call, we compute the petal decomposition of the graph. As explained in the previous subsection, this returns a set of petals $X_j$, their centers $x_j$, a set of neighbors $y_j$ and a set of targets $t_j$. We then run hierarchical-petal-decomposition recursively on each petal $X_j$ where the starting point is $x_j$ and the target is $t_j.$ Each of these recursive calls generates a tree $T_j$ for each petal $X_j.$ These trees are all disjoint, so we create a new graph which combines all of these $T_j$ through the edges $(x_j, y_j).$ This gives us a final low-stretch spanning tree $T.$

Lastly, to compute the tree itself, we call hierarchical-petal-decomposition on the graph with arbitrary starting point $x_0$ but critically also with target $x_0.$ The justification for this decision is explained in \cite{AN19}.

\subsubsection{LSST Implementation Validation}

The function hierarchical\_petal\_decomposition was the last function described in \cite{AN19}. In order to test the LSST implementation, we generate many Erd\H{o}s-R\'{e}nyi random graphs. For each random graph, we pick a probability $p$ and let there be an edge between every pair of distinct vertices with probability $p$ (so $p = 1$ would correspond to a complete graph). 

Then, for each graph, we compute an LSST using hierarchical\_petal\_decomposition, measuring its stretch and the duration to compute it. We repeat this for 10 trials and take the average for both the stretch and the duration. There are two primary things to verify: that the stretch is $O(m \log n \log \log n)$ and that the time complexity is $O(m \log n \log \log n).$ We also separately checked that the output was always a spanning tree (it reaches every vertex of the original graph and has $n - 1$ edges). However, when timing the code, we did not include this extra check since it is not a necessary part of the algorithm once we are confident that the algorithm is running correctly.

To check the asymptotic nature of the stretch and duration to compute the LSST, for different values of $m, n$ we graph the stretch and duration vs $m \log n \log \log n.$ We expect that there are some finite constants $c_1, c_2 \in \mathbb R$ such that the stretch with $m$ edges and $n$ vertices is bounded above by $c_1 m \log n \log \log n$ and likewise the duration is bounded above by $c_2 m \log n \log \log n.$ We check this graphically below.

\begin{figure}[H]
    \centering
    \includegraphics[scale = 0.9]{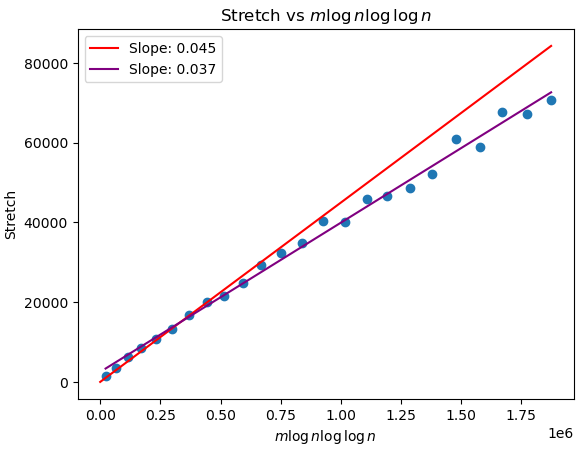}
    \caption{For different random graphs, we graph the average stretch vs $m \log n \log \log n.$ Empirically, we can see that the stretch is bounded above by $0.045 m \log n \log \log n$ and that the pattern of growth appears to suggest that this growth rate will continue. The purple line is the best fit line which minimizes squared L2 error.}
\end{figure}

\begin{figure}[H]
    \centering
    \includegraphics[scale = 0.9]{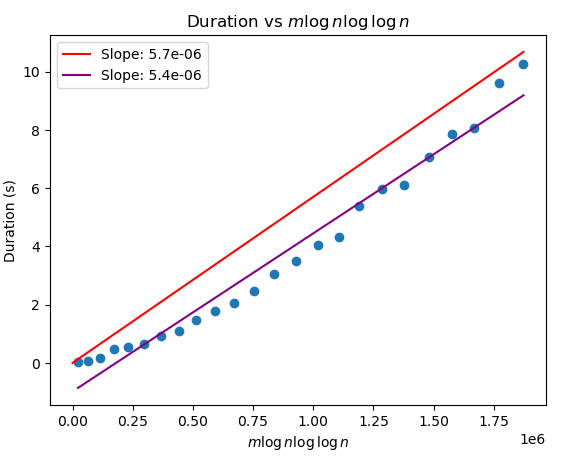}
    \caption{For different random graphs, we graph the average duration vs $m \log n \log \log n.$ Empirically, we can see that the duration is bounded above by $5.7 \cdot 10^{-6} m \log n \log \log n$ and that the pattern of growth appears to suggest that this growth rate will continue. The purple line is the best fit line which minimizes squared L2 error.}
\end{figure}

In both cases, we can see that for sufficiently large values of $m$ and $n,$ both the duration and stretch are bounded above by a linear constant multiplied by $m \log n \log \log n.$ Of course, this does not definitively verify that the implementation is correct (for example, there could hypothetically be a $\log \log \log n$ factor that would not show up graphically in the experiments run above) but given that the authors in \cite{AN19} prove their algorithm and stretch are both $O(m\log n \log \log n)$ and the implementation followed their algorithm, it is reasonable to conclude that the graphs above provide empirical validation that our implementation works correctly. 

\subsection{Heavy Light Decomposition}\label{heavy_light_section}

In this subsection we discuss the code that implements the Heavy Light Decomposition of a tree. First, we reproduce a necessary definition from \cite{main}, which we have modified slightly to refer to general trees:

\begin{defn} (\cite{main}) \textit{Given a rooted tree $T$ and vertex $u,$ define $u^{\uparrow T}$ as the set of its ancestors in $T$ plus itself. We extend the notation to any subset of vertices by defining $R^{\uparrow T} = \bigcup_{u \in R} u^{\uparrow T}.$}
    
\end{defn}

Now we define the Heavy Light Decomposition as described in Lemma B.8 of \cite{main} (which quotes \cite{st83}) which we reproduce below for convenience: 

\begin{lem}\label{heavy_light} 
There is a linear-time algorithm that given a rooted tree $T$ with $n$ vertices outputs a collection of vertex disjoint tree paths $\{P_1, \ldots, P_t\}$ (called heavy chains), such that the following hold for every vertex $u$:

\begin{enumerate}
    \item There is exactly one heavy chain $P_i$ containing $u.$
    \item If $P_i$ is the heavy chain containing $u$, at most one child of $u$ is in $P_i.$
    \item There are at most $O(\log n)$ heavy chains that intersect with $u^{\uparrow T}.$
\end{enumerate}

In addition, edges that are not covered by any heavy chain are called light edges.

\end{lem}

\subsubsection{Implementation Details}

The heavy\_light\_decomposition function takes in one input: the tree. Within this function, we define two inner functions: dfs\_size and decompose\_chain, which we describe below.

For the dfs\_size function which takes in a vertex $v$ and its parent as inputs. Then, the function uses Depth First Search to recursively visit every vertex and compute the size of the subtree (number of vertices) rooted at that vertex. If the vertex is a leaf node, we say its subtree has size 1 (just itself). Otherwise, we initialize the size of the subtree to 1 and compute $\size(v) = 1 + \sum_{w \in c(v)} \size(w)$ where $c(v)$ is the set of all children of $v$. 

Next, we have the decompose\_chain function. This function takes in a vertex $v$, the head of the chain of $v$ and the parent of $v$ as inputs. In this function, we first identify the heavy child for that vertex, which corresponds to condition 2 in Definition \ref{heavy_light} and is the one child of $u$ within $P_i$, the heavy chain containing $u$. We find this heavy child by iterating over the children of $u$ and picking the child of $u$ which has the maximum subtree size. 

The decompose\_chain function then recursively calls itself as long as we keep finding new ``heavy children." For each recursive call, we take in the previous heavy\_child as our new vertex $v,$ the same chain\_head as before and the vertex $v$ as the parent (since the heavy\_child is a child of vertex $v$). 

Once the recursive calls are complete, we iterate over the children of $v$, and for every child that isn't the heavy child, we initialize a new chain and recursively call decompose\_chain where our new vertex $v$ is the child we're iterating over, the chain\_head is the index of the current chain and the parent is our old $v.$ By setting the code up in this way, we ensure that we create a chain for every vertex $v$ and its descendants and that we satisfy the requirements of Lemma \ref{heavy_light}.

With these two functions implemented, we can finish our heavy\_light\_decomposition implementation. We initialize variables for the length of the tree, the subtree sizes, and the chains. We then call dfs\_size on the root vertex $0$ and its ``parent" (which is not a real vertex) but which we denote $-1.$ This allows us to figure out the sizes of all of the subtrees in the tree. Then, we call decompose\_chain on our root vertex $0$, the chain\_head which we initialize to $0$ and the parent $-1$. Finally, the heavy\_light decomposition function returns the list of chains.

\subsubsection{Heavy-Light Decomposition Implementation Validation}

In order to check our implementation, we check the three conditions of Lemma \ref{heavy_light} empirically. We produce a function that generates random trees. These trees are implemented as a list of lists, with nodes numbered $0$ to $n$ where $0$ is the root. Further, the list at index $i$ consists of the children of node $i$, and the list is empty if and only if node $i$ is a leaf node. Based on this format, such random trees can be generated as follows: for vertex $n$, randomly choose one of the vertices $i = 0, 1, \ldots, n - 1$ to be its parent by appending vertex $n$ to the list at index $i$. This also has the convenient property that the parent's number is always strictly less than the child's number.

We then call our heavy light decomposition function on 1000 randomly generated trees. We iterate over all possible vertices for that tree and check that vertex $u$ is represented in one and only one heavy chain. We can then easily check the children of $u$ and ensure that at most one child of $u$ is in that same heavy chain. 

For the third condition, we calculate $u^{\uparrow T}$ for each $u$ and count the number of heavy chains that intersect with it. This is the trickiest condition to check because Lemma \ref{heavy_light} only promises an $O(\log n)$ bound on the number of intersecting heavy chains. To check this, we consider increasing values of $n$ and graph the average number of intersecting heavy chains vs $\log n$, where the average is calculated by summing up the number of intersections across all vertices and then dividing by both the number of vertices and the number of randomly generated trees, in this case 1000. Below, we plot the average number of intersections vs $\log n$, and we see that a line with slope 1.15 dominates the growth. Thus, empirically it appears that the number of intersections is bounded by $1.15 \log n.$ In addition, we also computed the quotient of the number of intersections divided by $\log n$, which gave us the following list: [1.276, 1.196, 1.144, 1.134, 1.126, 1.103, 1.092, 1.090, 1.084,
       1.081, 1.079, 1.072]. We can see that this list is monotonically decreasing, which provides further assurance that the number of intersections is $O(\log n).$

\begin{figure}[H]
    \centering
    \includegraphics[scale = 0.9]{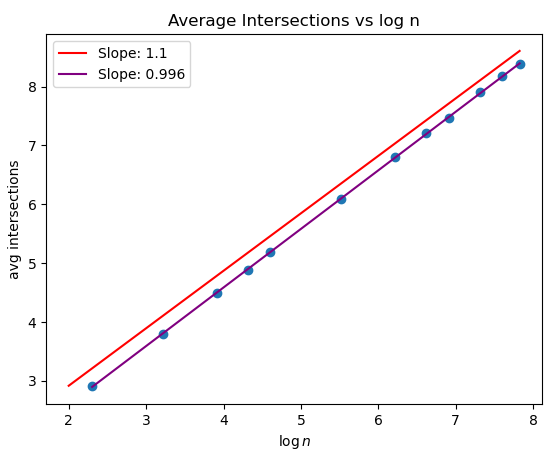}
    \caption{The average number of intersections appears to be bounded by $O(\log n)$. The purple line is the best fit line.}
    \label{fig:enter-label}
\end{figure}

In addition, we also empirically check the time complexity for our implementation just like we did for low-stretch spanning trees. Lemma 4.4 states that the algorithm should be $O(n)$ time, which indeed appears to be the case below.

\begin{figure}[H]
    \centering
    \includegraphics[scale = 0.9]{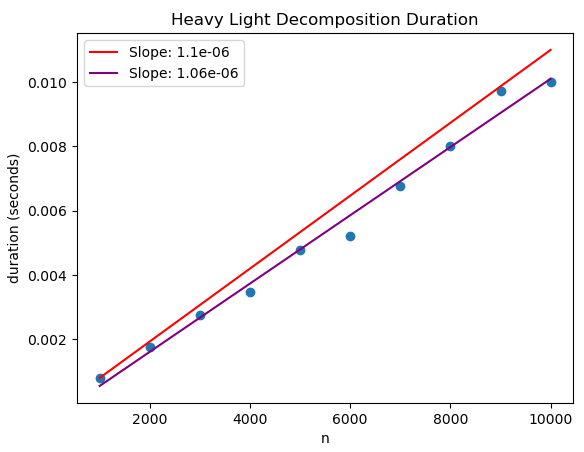}
    \caption{The time to compute the heavy light decomposition increases linearly with the number of vertices, which matches our desired $O(n)$ time complexity. The purple line is the best fit line.}
    \label{fig:enter-label}
\end{figure}

\section{Rebuilding Game}\label{rebuilding_section} 

The authors in \cite{main} create a \textit{rebuilding game} in order to attain their promised near linear time bound for max flow. As they describe at the beginning of section 8, this rebuilding game is the key step to go from Theorem 7.1 to Theorem 6.2 (theorem numbers are with respect to \cite{main}). Theorem 6.2 asserts the existence of the main overall data structure in \cite{main} that is crucial for attaining the near asymptotic linear time bound. The code for this part is included in the file titled rebuilding.ipynb. Although for our purposes we only need the rebuilding game under very specific conditions with specific parameter values, we implement it in full generality as it is described in Section 8 of \cite{main}.

\subsection{Preliminaries}

``The rebuilding game has several parameters: integers parameters size $m > 0$ and depth $d > 0$, update frequency $0 < \gamma_g < 1$, a rebuilding cost $C_r \geq 1$, a weight range $K \geq 1$, and a recursive size reduction parameter $k = m^{1/d} \geq 2$, and finally an integer round count $T > 0$" (\cite{main}).

They define the rebuilding game in Definition 8.1 of \cite{main} which we reproduce here for convenience:

\begin{defn} (\cite{main}) \textit{The rebuilding game is played between a player and an adversary and proceeds in rounds $t = 1, 2, \ldots, T.$ Additionally, the steps (moves) taken by the player are indexed as $s = 1, 2, \ldots$. Every step $s$ is associated with a tuple $\text{prev}^{(s)} := (\text{prev}_0^{(s)}, \ldots, \text{prev}_d^{(s)}) \in [T]^{d + 1}.$ Both the player and adversary know $\text{prev}^{(s)}$. At the beginning of the game, at round $t = 1$ and step $s = 1$, we initially set $\text{prev}_i^{(1)} = 1$ for all levels $i \in \{0, 1, \ldots, d\}.$ At the beginning each round $t \geq 1$,}

\begin{enumerate}\label{rebuilding_defn}
    \item \textit{The adversary first chooses a positive real weight $W^{(t)}$ satisfying $\log W^{(t)} \in (-K, K).$ This weight is \textbf{hidden} from the player.}

    \item \textit{Then, \textbf{while} either of the following conditions hold, $\sum_{i = 0}^d W^{\text{prev}_i^{(s)}} > 2(d + 1)W^{(t)}$ or for some level $l,$ at least $\gamma_g m/k^l$ rounds have passed since the last rebuild of level $l,$ the adversary can (but does not have to) force the player to perform a fixing step. The player may also choose to perform a fixing step, regardless of whether the adversary forces it or not. In a fixing step, the player picks a level $i \in \{0, 1, \ldots, d \}$ and we then set $\text{prev}_j^{(s + 1)} \leftarrow t$ for $j \in \{i, i + 1, \ldots, d\},$ and $\text{prev}_j^{(s + 1)} \leftarrow \text{prev}_j^{(s+1)}$ for $j \in \{0, \ldots, i - 1\}.$ We call this a fix at level $i$ and we say the levels $j \geq i$ have been rebuilt. This move costs $C_r m/k^i$ time.}

    \item \textit{When the player is no longer performing fixing steps, the round finishes.}
\end{enumerate}

\end{defn}

There are some aspects of the original explanation in \cite{main} that were underspecified and are clarified here. The first is that the steps indexed by $s$ refer to fixing steps, and further that each step also has a level $i$. For this reason, the step count is incremented each time a level is fixed (thus $s$ refers to the total number of fixing steps across all levels). Note that this increment of the step count is separate from the fixing counts in the fix array, which are incremented as well until they are reset. 


In order to assess whether the adversary is forcing a fixing step, we check the $\sum_{i = 0}^d W^{\text{prev}_i^{(s)}} > 2(d + 1)W^{(t)}$ as specified above in Definition \ref{rebuilding_defn}.

\subsection{Implementation and Validation}

We implement this rebuilding game through two functions. We first implement a fix function which the main algorithm will call to perform a fixing step. This function is fairly routine and simply takes in a level $i$, an index $j$, the step index $s$, the round $t$ and an array called prevs. The fix function returns the newly updated prevs array after the fix is complete.

Next, we have our main rebuilding function. In addition to the parameters $m, d, \gamma_g, C_r, K, T$ listed above, we also take in the parameter $w,$ which represents a 2d array where the $t^{th}$ row of $w$ is $w^{(t)}$ and we initialize $W^{(t)} = ||w^{(t)}||_1$ as explained in Remark 8.2 of \cite{main}. We also initialize 1D arrays to keep track of the fix and round counts and a 2D prevs array. The prevs array has $d$ columns (one for each level) and initially one row of all ones as outlined in Definition \ref{rebuilding_defn} above.

We provide empirical validation that the algorithm runs in $O\left(\frac{C_r K d}{\gamma_g}(m + T) \right).$ For setting the values of each of these variables, we refer to Section 8.2 of \cite{main}, which gives asymptotic values for the variables. We let $C_r = \exp(\log^{7/8} m \log \log m), \gamma_g = \frac{1}{C_r} = \exp(-\log^{7/8} m \log \log m), Q = mC_r = m \exp(\log^{7/8} m \log \log m), T = \lfloor (m + Q) C_r \rfloor$ and $d = \lfloor 5 \cdot \sqrt[8]{\log m} \rfloor.$ While section 8.2 of \cite{main} states that $d = \sqrt[8]{\log m}$ as opposed to $O(\sqrt[8]{\log m}),$ we note that scaling $d$ by a constant has no bearing on the $O\left(\frac{C_r K d}{\gamma_g}(m + T) \right)$ runtime, and the authors also state $d = O(\sqrt[8]{\log m})$ many other times in \cite{main}. The reason for the constant factor of $5$ in this implementation is because otherwise we would always have $d = 1$ unless $m$ was extremely large, and there is no reason to always set $d = 1.$ The exact value of $5$ is arbitrary and can be replaced by any other positive integer.

In addition, we also let $W$ be a vector with $T$ entries where the $W[i]$ are i.i.d Uniform(1, 1000). Here, the number 1000 is arbitrary. Lastly, we set $K = \lceil \log \max_i W[i] \rceil$ in order to ensure that $\log W[i] \in (-K, K)$ as required by \cite{main}.

With these parameters, we run experiments with $m = 8, 9, 10 \ldots, 64.$ We plot the durations for each of these values of $m$ vs $\frac{C_r K d}{\gamma_g}(m + T).$ As promised by \cite{main}, we do empirically observe that the runtime increases linearly with $\frac{C_r K d}{\gamma_g}(m + T)$ and is therefore bounded by a constant times $\frac{C_r K d}{\gamma_g}(m + T),$ where this constant is the slope of the red line in the graph below or $2.02 \cdot 10^{-11}.$ This number comes from calculating the slope between the last points and one of the preceding points and increasing the y-intercept slightly, as the graph appears to be progressing linearly once the $x$ axis exceeds $4.$

\begin{figure}[H]
    \centering
    \includegraphics[scale = 0.8]{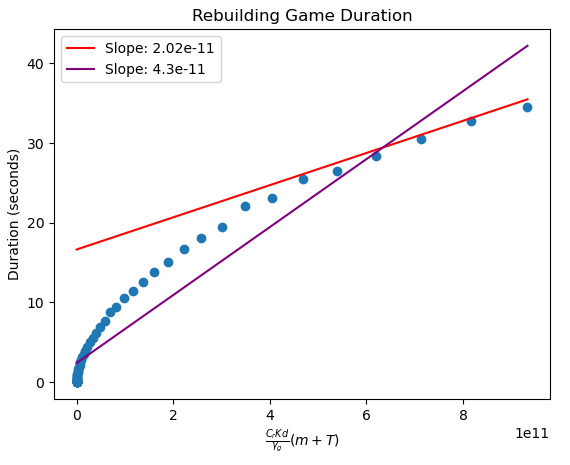}
    \caption{The runtime of the rebuilding game appears to be $O\left(\frac{C_r K d}{\gamma_g}(m + T)\right)$ as stated in \cite{main}. The purple line is the best fit line.}
    \label{fig:enter-label}
\end{figure}

\section{Interior Point Method}\label{ipm_section}

In Section 4 of \cite{main}, the authors present an interior point method that solves the min-cost flow problem. 

\subsection{IPM Preliminaries}

We begin by defining the key variables that will be used throughout this section and the remainder of the paper as they are defined in \cite{main}. In the subsequent text, recall that $\mathbf{B}$ was defined in Notation \ref{B_defn}.

\begin{defn}\label{vars_defn} For our min-cost flow problem, we let $\mathbf{d} \in \mathbb Z^V$ represent the demands. We also let the lower and upper capacities be denoted by $\mathbf{u}^{-}, \mathbf{u}^+ \in \mathbb Z^E$ where all integers are bounded by the variable $U$. Finally, we let $\mathbf{c} \in \mathbb Z^E$ denote the costs.
    
\end{defn}

\begin{defn} We define $\alpha \stackrel{\text{def}}{=} \frac{1}{1000 \log mU}.$ We will use $\alpha$ to create a barrier $x^{-\alpha}$ when solving the L1 IPM.
\end{defn}

\begin{defn}\label{f_star} We define:

$$\mathbf{f}^* \stackrel{\text{def}}{=} \argm_{\substack{\mathbf{B^\top \mathbf{f}} = d \\ \mathbf{u}_e^- \leq \mathbf{f}_e \leq \mathbf{u}_e^+ \text{for all} e \in E}} c^T f.$$
    
\end{defn}

\begin{defn}\label{defn_potential} We let $\Phi(\mathbf{f}) \stackrel{\text{def}}{=} 20m \log(c^\top \mathbf{f} - F^*) + \sum_{e \in E} ((\mathbf{u}_e^+ - \mathbf{f}_e)^{-\alpha} + (\mathbf{f}_e - \mathbf{u}_e^{-})^{-\alpha}).$
    
\end{defn}
    
\begin{defn}\label{defn4.2} (\cite{main}) Given a flow $\mathbf{f} \in \mathbb R^E$ we define lengths $\ell \in \mathbb R^E$ as

$$\mathbf{\ell}(\mathbf{f})_e \stackrel{\text{def}}{=} (\mathbf{u}_e^+ - \mathbf{f}_e)^{-1 - \alpha} + (\mathbf{f}_e - \mathbf{u}_e^-)^{-1 - \alpha}$$

and gradients $\mathbf{g} \in \mathbb R^E$ as $\mathbf{g}(\mathbf{f}) \stackrel{\text{def}}{=} \nabla \Phi(\mathbf{f}).$ More explicitly,

$$\mathbf{g}(\mathbf{f})_e \stackrel{\text{def}}{=} [\nabla \Phi(\mathbf{f})]_e = 20m(\mathbf{c}^\top \mathbf{f} - F^*)^{-1} \mathbf{c}_e + \alpha(\mathbf{u}_e^+ - \mathbf{f}_e)^{-1 - \alpha} - \alpha(\mathbf{f}_e - \mathbf{u}_e)^{-1 - \alpha}.$$
    
\end{defn}

We now reproduce Theorem 4.3 of \cite{main}, which is their main result regarding IPMs for solving the min-cost flow problem ($f^{(t)}$ is the flow after iteration $t$, while the optimal flow $f^*$ is unknown):

\begin{thm}\label{thm4.3} (\cite{main}) Suppose we are given a min-cost flow instance given by Equation (1). Let $\mathbf{f}^*$ denote an optimal solution to the instance.

For all $\kappa \in (0, 1),$ there is a potential reduction interior point method for this problem, that, given an initial flow $\mathbf{f}^{(0)} \in \mathbb R^E$ such that $\Phi(\mathbf{f}^{(0)}) \leq 200m \log mU,$ the algorithm proceeds as follows:

The algorithm runs for $\tilde O(m\kappa^2)$ iterations. At each iteration, let $\mathbf{g(f^{(t)})} \in \mathbb R^E$ denote that gradient and $\mathbf{\ell}(\mathbf{f}^{(t)}) \in \mathbb R^E_{> 0}$ denote the lengths given by Definition \ref{defn4.2}. Let $\mathbf{\tilde g} \in \mathbb R^E$ and $\mathbf{\tilde \ell} \in \mathbb R_{>0}^E$ be any vectors such that $||\mathbf{L(f^{(t)}})^{-1}(\mathbf{\tilde g} - \mathbf{g}(f^{(t)}))||_{\infty} \leq \kappa/8$ and $\mathbf{\tilde \ell} \approx \mathbf{\ell}(f).$

\begin{enumerate}
    \item At each iteration, the hidden circulation $\mathbf{f}^* - \mathbf{f}^{(t)}$ satisfies
    $$\frac{\mathbf{g}(\mathbf{f}^* - \mathbf{f}^{(t)})}{100m + ||\tilde{\mathbf{L}}(\mathbf{f}^* - \mathbf{f}^{(t)})||_1} \leq -\alpha/4.$$
    \item At each iteration, given any $\Delta$ satisfying $\mathbf{B}^\top \delta = 0$ and $\tilde{\mathbf{g}}\Delta/||\tilde{\mathbf{L}} \Delta||_1 \leq -\kappa,$ it updates $\mathbf{f}^{(t + 1)} \leftarrow \mathbf{f}^{(t)} + \eta \Delta$ for $\eta \leftarrow \kappa^2/(50 \cdot |\tilde{\mathbf{g}} \Delta|).$
    \item At the end of $\tilde O(m \kappa^2)$ iterations, we have $\mathbf{c}^\top \mathbf{f}^{(t)} \leq \mathbf{c}^\top \mathbf{f}^{*} + (mU)^{-10}.$
\end{enumerate}
    
\end{thm}

\subsection{IPM Implementation}\label{ipm_implementation}

All code described in this section can be found in the file ipm.ipynb. 

In \cite{main}, the authors give a fairly detailed implementation that relies on specific  bounds on the gradient, number of iterations etc. In order to implement that version of an interior point method, we would have to implement an oracle for a specific step described in Section 4 of \cite{main}, which is nontrivial to implement.

In this section we merely provide an approximation/heuristic for solving the interior point method which relies on scipy.optimize. We define the objective function as given in Definition \ref{defn_potential} and feed the constraints as well. The minimize function in scipy.optimize then numerically solves for the optima of the potential function.

We empirically check the runtime of this simple implementation and compare it to the $\tilde O(m \kappa^2)$ runtime promised above by Theorem \ref{thm4.3}, where we let $\kappa^2 = \exp(-\log^{7/8} m \log \log m)$ as defined later on in \cite{main} (they say $\kappa = \exp(-O(\log^{7/8} m \log \log m)),$ but an extra constant factor doesn't change the following analysis). 

\begin{figure}[H]
    \centering
    \includegraphics[scale = 0.9]{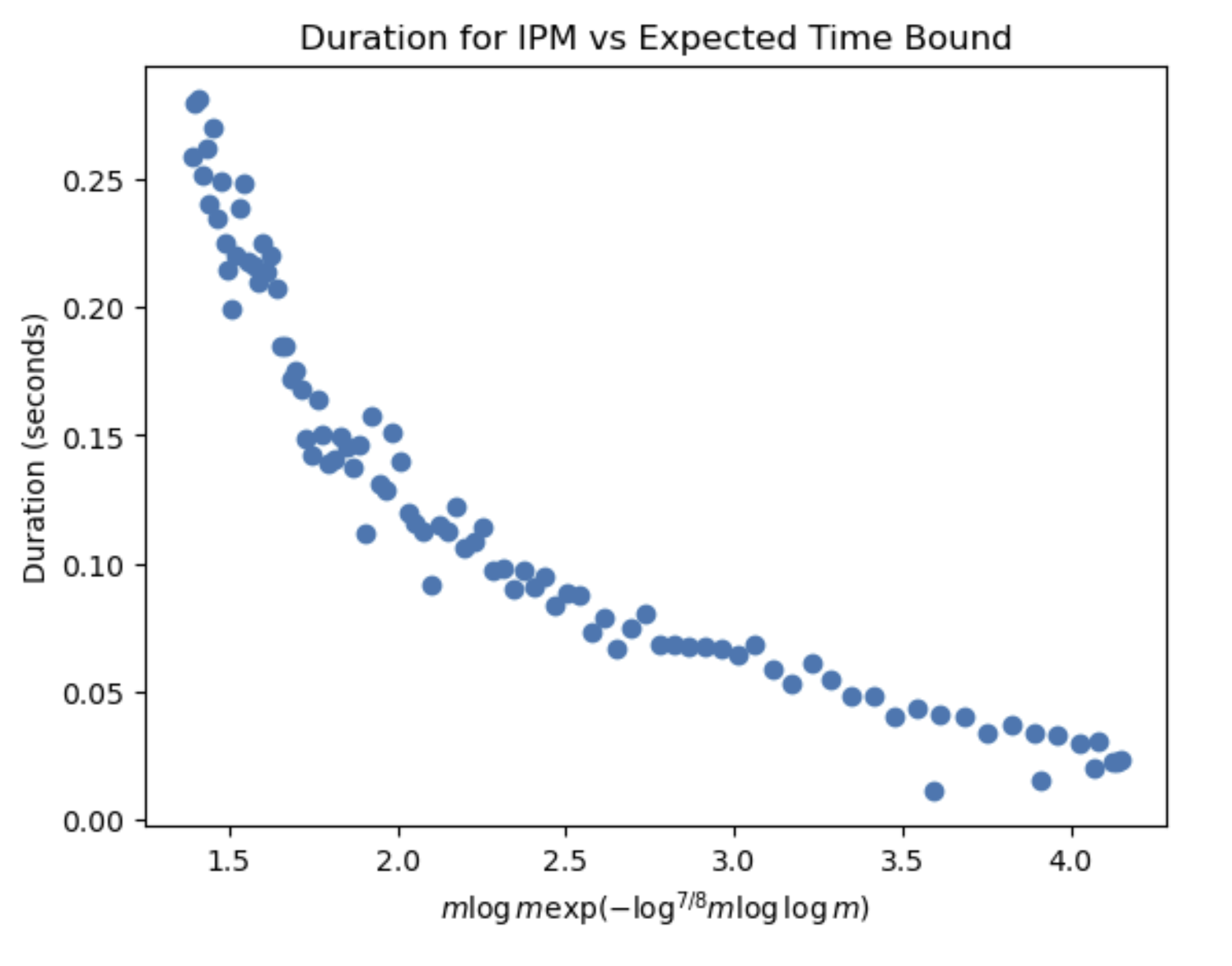}
    \label{fig:enter-label}
\end{figure}

We notice that the graph is decreasing, which is surprising. Empirically checking the values of $m \log m \exp(-\log^{7/8} m \log \log m)$, we find they are decreasing over the values from $m = 4$ to $m = 100.$ In order to inspect this, we take the log of this expression, where we see:

$$\lim_{m \to \infty} \log m + \log \log m - \log^{7/8} m \log \log m = \lim_{m \to \infty} \log m\left(1 + \frac{\log \log m}{\log m} - \frac{\log \log m}{\sqrt[8]{\log m}} \right) = \infty.$$

Specifically, in the final expression above, we know that $\log m$ will go to $\infty$ while the expression in the parentheses will go to $1$. Thus in the limit the log goes to infinity, which means the original expression goes to infinity as well. Yet the empirically decreasing nature of the function in the graph above raises a question: when does this function start increasing?

To find this, we take a derivative of the logged expression. We first let $x = \log m$ since $m$ only appears as $\log m.$ Then, we have:

$$\frac{d}{dx} (x + \log x - x^{7/8} \log x) = \frac{1}{x} + 1 - \frac{7 \log x}{8\sqrt[8]{x}} - \frac{1}{\sqrt[8]{x}}.$$

We want to find out when the derivative is positive. If we take $x \to \infty$, we see that the derivative approaches $1,$ but if we plug in a number like $x = 1000$ (which, we keep in mind corresponds to $m = e^{1000}$) we get $-1.97.$ This function is continuous, so by the Intermediate Value Theorem, we know that it will cross $0$ at some point. We will find out where the derivative equals $0$ (which corresponds to a local optimum) and then argue that it increases after this point.

Setting our above expression equal to $0$, we get:

$$\frac{1}{x} + 1 = \frac{7 \log x + 8}{8 \sqrt[8]{x}} \implies 7 \log x + 8 = 8x^{1/8} - \frac{8}{x^{7/8}}.$$

If we plug this expression into Wolfram Alpha, we get $x \approx 7631410337,$ which we will write as $x = 7.63 \cdot 10^{10}$ (but all numerical calculations below are done using the more precise value of $x$). Now, we want to confirm that the derivative is positive whenever $x > 7.63 \cdot 10^{10}.$ We have:

$$\frac{1}{x} + 1 - \frac{7 \log x}{8\sqrt[8]{x}} - \frac{1}{\sqrt[8]{x}} > 0 \iff \frac{8}{x^{7/8}} + 8x^{1/8} > 7 \log x + 8.$$

The derivative of the left hand side is $\frac{-7}{x^{15/8}} + x^{-7/8}$ while the derivative of the right hand side is $\frac{7}{x}.$ Then, we see:

$$\frac{1}{x^{7/8}} - \frac{7}{x^{15/8}} > \frac{7}{x} \iff x^{1/8} - \frac{7}{x^{7/8}} > 7.$$

From Wolfram Alpha, we see this is true whenever $x > 5764860$ (which makes sense as $7^8 = 5764801$) and $7.63 \cdot 10^{10} > 5764860.$ This means when $x = 7.63 \cdot 10^{10},$ $\frac{8}{x^{7/8}} + 8x^{1/8}$ is increasing faster than $7 \log x + 8,$ so the first derivative will be positive for $x > 7.63 \cdot 10^{10}$ and $x = 7.63 \cdot 10^{10}$ is a local minimum of the differentiated expression.

Returning to our original expression, we recall that $x = \log m,$ so our original function is increasing when $m > \exp(7.63 \cdot 10^{10}) \approx 10^{33142793985} \approx 10^{3.31 \cdot 10^{10}}.$ When $m = 10^{3.31 \cdot 10^{10}},$ we have $\log m = 7.63 \cdot 10^{10}, \log \log m = 25.06,  \sqrt[8]{\log m} = 22.93.$. Plugging these values in, at this local minimum, we have:

$$m \log m \exp(-\log^{7/8} m \log \log m) = m^{1 + \frac{\log \log m}{\log m} - \frac{\log \log m}{\sqrt[8]{\log m}}} = 10^{3.31 \cdot 10^{10}\left(1 + \frac{25.06}{7.63 \cdot 10^{10}} - \frac{25.06}{22.93}\right)}.$$

As stated previously, all calculations are done using the precise values of each numerical quantity rather than the rounded values we use above.

Simplifying, this comes out to $10^{-1.07 \cdot 10^{10}}$ which is of course extremely, extremely close to $0$. Thus, even though our function eventually tends to infinity, it first approaches $0$. Further, given that $m = 10^{3.31 \cdot 10^{10}}$ is many orders of magnitude larger than the number of atoms in the known universe (approximately $10^{82}$), it is not possible to empirically check whether our scipy implementation or (or even a fully correct implementation) matches the asymptotic growth of $m \log m \exp(-\log^{7/8} m \log \log m).$

For this reason, it does not make sense practically to consider the runtime of the IPM relative to its asymptotic guarantee. For any future implementations of the IPM algorithm described by Theorem \ref{thm4.3}, the focus should be on the other requirements of Theorem \ref{thm4.3}. With that said, it's hard to test the other requirements either. While we could calculate $f^*$ via another algorithm to try to test our IPM implementation for accuracy, the scipy.optimize method that we are using only returns the final converged value rather than the successive steps which makes requirements 1 and 2 of Theorem \ref{thm4.3} infeasible to check. Even if we had the successive step values, it is unlikely that this heuristic would exactly match those requirements. For requirement 3, we note that the $(mU)^{-10}$ bound only holds  
after requires $\tilde O(m \kappa^{-2})$ iterations which could be extremely large practically, which suggests that the scipy implementation may run into numerical issues or may not converge in time, which makes it unreliable for testing. Thus, while the scipy implementation provides a heuristic for solving the IPM problem, it is not complete.

\section{Overall Time Complexity}\label{overall}

The authors in \cite{main} state that their max flow algorithm is $m^{1 + o(1)} \log U \log C$ time or what they sometimes call $\tilde O(m)$ time. The goal of this section is to more precisely calculate the $m^{o(1)}$ term.

For a graph with $m$ edges and $n$ vertices to be connected, we must have at least $n - 1$ edges. Therefore, $n - 1 \leq m$ which implies that $n = O(m).$ This allows us to bound $n$ by $m$ in the asymptotic calculations that follow. 

We refer to Algorithm 7 of \cite{main}, which provides the algorithm for their main theorem. We then refer to each component of it and calculate its runtime based on what the authors in \cite{main} say. In cases when they simply say $m^{o(1)},$ we take a deeper dive as needed to give a more precise asymptotic time complexity.

\subsection{Preliminaries}

We begin by reproducing several Lemmas and Theorems from \cite{main} which we will refer to throughout the subsequent subsections:

\begin{defn}\label{defn6.1} (\cite{main}) \textit{ Consider a dynamic graph $G^{(t)}$ undergoing batches of updates $U^{(1)}, \ldots, U^{(t)}, \ldots$ consisting of edge insertions/deletions and vertex splits. We say the sequences $\mathbf{g}^{(t)}, \mathbf{\ell}^{(t)},$ and $U^{(t)}$ satisfy the \textbf{hidden stable-flow chasing} property if there are hidden dynamic circulations $\mathbf{c}^{(t)}$ and hidden dynamic upper bounds $\mathbf{w}^{(t)}$ such that the following holds at all stages $t$:}

\begin{enumerate}
    \item \textit{$\mathbf{c}^{(t)}$ is a circulation: $\mathbf{B}^{\top}_{G^{(t)}} \mathbf{c}^{(t)} = 0.$}
    \item \textit{$\mathbf{w}^{(t)}$ upper bounds the length of $\mathbf{c}^{(t)}$: $|\mathbf{\ell}_e^{(t)} \mathbf{c}_e^{(t)}| \leq \mathbf{w}_e^{(t)}$ for all $e \in E(G^{(t)}).$}
    \item \textit{For any edge $e$ in the current graph $G^{(t)}$, and any stage $t' \leq t,$ if the edge $e$ was already present in $G^{t'},$ i.e. $e \in G^{(t)}\backslash \bigcup_{s = t' + 1}^t U^{(s)}$ then $\mathbf{w}_e^{(t)} \leq 2\mathbf{w}_e^{(t')}.$}
    \item \textit{Each entry of $\mathbf{w}^{(t)}$ and $\mathbf{\ell}^{(t)}$ is quasipolynomially lower and upper-bounded: $\log \mathbf{w}_e^{(t)} \in [-\log^{O(1)} m, \log^{O(1)} m]$ and $\log \mathbf{\ell}_e^{(t)} \in [-\log^{O(1)} m, \log^{O(1)} m]$ for all $e \in E(G^{(t)}).$}
\end{enumerate}
    
\end{defn}

\begin{defn}\label{defn6.7} (\cite{main}) Consider a tree $T$ and a rooted spanning forest $E(F) \subseteq E(T)$ on a graph $G$ equipped with stretch overestimates $\widetilde{str}_e$ satisfying the guarantees of Lemma \ref{lem6.5}. We define the core graph $\mathcal{C}(G, F)$ as a graph with the same edge and vertex set as $G \backslash F.$ For $e = (u, v) \in E(G)$ with image $\hat{e} \in E(G/F)$ we define its length as $\mathbf{\ell}_{\hat{e}}^{\mathcal{C}(G, F)} \stackrel{\text{def}}{=} \widetilde{str}_e \mathbf{\ell}_e$ and gradient as $\mathbf{g}_{\hat{e}}^{\mathcal{C}(G, F)} \stackrel{\text{def}}{=} \mathbf{g}_e + \langle \mathbf{g}, \mathbf{p}(T[u, v]) \rangle.$
    
\end{defn}

\begin{defn}\label{defn6.9} (\cite{main}) Given a graph $G$, forest $F$, and parameter $k$, define a $(\gamma_s, \gamma_c, \gamma_\ell)$-sparsified core graph with embedding as a subgraph $\mathcal{S}(G, F) \subseteq \mathcal{C}(G, F)$ and embedding $\prod_{\mathcal{C}(G, F) \to \mathcal{S}(G, F)}$ satisfying 

\begin{enumerate}
    \item For any $\hat{e} \in E(\mathcal{C}(G, F))$, all edges $\hat{e}' \in \prod_{\mathcal{C}(G, F) \to \mathcal{S}(G, F)}(\hat{e})$ satisfy $\mathbf{\ell}_{\hat{e}}^{\mathcal{C}(G, F)} \approx_2 \mathbf{\ell}_{\hat{e}'}^{\mathcal{C}(G, F)}.$
    \item $\length(\prod_{\mathcal{C}(G, F) \to \mathcal{S}(G, F)}) \leq \gamma_\ell$ and $\econg(\prod_{\mathcal{C}(G, F) \to \mathcal{S}(G, F)}) \leq k \gamma_c.$
    \item $\mathcal{S}(G, F)$ has at most $m\gamma_s/k$ edges.
    \item The lengths and gradients of edges in $\mathcal{S}(G, F)$ are the same in $\mathcal{C}(G, F)$ (Definition \ref{defn6.7}).
\end{enumerate}
    
\end{defn}

\begin{defn}\label{defn6.10} (\cite{main}) For a graph $G,$ parameter $k,$ and branching factor $B,$ a \textbf{B-branching tree-chain} consists of collections of graphs $\{\mathcal{G}_i \}_{0 \leq i \leq d}$ such that $\mathcal{G}_0 \stackrel{\text{def}}{=} \{G\},$ and we define $\mathcal{G}_i$ inductively as follows,

\begin{enumerate}
    \item For each $G \in \mathcal{G}_i, i < d$, we have a collection of $B$ trees $\mathcal{T}^{G_i} = \{T_1, T_2, \ldots, T_B \}$ and a collection of $B$ forests $\mathcal{F}^{G_i} = \{F_1, F_2, \ldots, F_B \}$ such that $E(F_j) \subseteq E(T_j)$ satisfy the conditions of Lemma \ref{lem6.5}.
    \item For each $G_i \in \mathcal{G}_i,$ and $F \in \mathcal{F}^{G_i},$ we maintain $(\gamma_s, \gamma_c, \gamma_l)$-sparsified core graphs and embeddings $\mathcal{S}(G_i, F)$ and $\Pi_{\mathcal{C}(G_i, F) \to \mathcal{S}(G_i, F)}.$

    \item We let $\mathcal{G}_{i + 1} \stackrel{\text{def}}{=} \{\mathcal{S}(G_i, F) : G_i \in \mathcal{G}_i, F \in F^{G_i} \}.$

    Finally, for each $G_d \in \mathcal{G}_d,$ we maintain a low-stretch tree $F.$

    We let \textbf{a tree chain} be a single sequence of graphs $G_0, G_1, \ldots, G_d$ such that $G_{i + 1}$ is the $(\gamma_s, \gamma_c, \gamma_l)$-sparsified core graph $\mathcal{S}(G_i, F_i)$ with embedding $\Pi_{\mathcal{C}(G_i, F_i) \to \mathcal{S}(G_i, F_i)}$ for some $F_i \in \mathcal{F}^{G_i}$ for $0 \leq i \leq d,$ and a low-stretch tree $F_d$ on $G_d.$
\end{enumerate}
    
\end{defn}

\begin{defn}\label{defn6.11} (\cite{main}) Given a graph $G$ and tree-chain $G_0, G_1, \ldots, G_d$ where $G_0 = G,$ define the corresponding spanning tree $T^{G_0, G_1, \ldots, G_d} \stackrel{\text{def}}{=} \bigcup_{i = 0}^d F_i$ of $G$ as the union of preimages of edges of $F_i$ in $G = G_0.$

Define the set of trees corresponding to a branching tree-chain of graph $G$ as the union of $T^{G_0, G_1, \ldots, G_d}$ over all tree chains $G_0, G_1, \ldots, G_d$ where $G_0 = G$:

$$\mathcal{T}^G \stackrel{\text{def}}{=} \{T^{G_0, G_1, \ldots, G_d} : G_0, G_1, \ldots, G_d ~s.t.~ G_{i + 1} = \mathcal{S}(G_i, F_i) \text{ for all } 0 \leq i < d \}$$
    
\end{defn}

\begin{defn}\label{defn6.12} (\cite{main}) Given a dynamic graph $G^{(t)}$ with updates indexed by times $t = 0, 1, \ldots$ and corresponding dynamic branching tree-chain (Definition \ref{defn6.10}), we say that nonnegative integers $\text{prev}_0^{(t)} \leq \text{prev}_1^{(t)} \leq \cdots \leq \text{prev}_d^{(t)}$ are previous rebuild times if $\text{prev}_i^{(t)}$ was the most recent time at or before $t$ such that $\mathcal{G}_i$ was rebuilt, i.e. $G \in \mathcal{G}_i$ the set of trees $\mathcal{T}^G$ was reinitialized and sampled.
    
\end{defn}

\begin{lem}\label{lem8.3} (\cite{main}) There is a deterministic strategy given by Algorithm 6 for the player to finish $T$ rounds of the rebuilding game in time $O\left(\frac{C_r Kd}{\gamma_g}(m + T) \right).$
    
\end{lem}

\begin{thm}\label{thm7.1} (\cite{main}) Let $G = (V, E)$ be a dynamic graph undergoing $\tau$ batches of updates $U^{(1)}, \ldots, U^{(\tau)}$ containing only edge insertions/deletions with edge gradient $\mathbf{g}^{(t)}$ and length $\mathbf{\ell}^{(t)}$ such that the update sequence satisfies the hidden stable-flow chasing property (Definition \ref{defn6.1}) with hidden dynamic circulation $\mathbf{c}^{(t)}$ and width $\mathbf{w}^{(t)}.$ There is an algorithm on $G$ that maintains a $O(\log n)-$ branching tree chain corresponding to $s = O(\log n)^d$ trees $T_1, T_2, \ldots, T_s$ (Definition \ref{defn6.11}), and at stage $t$ outputs a circulation $\Delta$ represented by $\exp(O(\log^{7/8} m \log \log m))$ off-tree edges and paths on some $T_i, i\in [s].$ 

The output circulation $\Delta$ satisfies $\mathbf{B}^\top \Delta = 0$ and for some $\kappa = \exp(-O(\log^{7/8} m \log \log m))$

$$\frac{\langle \mathbf{g}^{(t)}, \Delta \rangle}{||\mathbf{\ell}^{(t)} \circ \Delta||_1} \leq \kappa \frac{\langle \mathbf{g}^{(t)}, \mathbf{c}^{(t)} \rangle}{\sum_{i = 0}^d ||\mathbf{w}^{(\text{prev}_i^{(t)})}||_1},$$

where $\text{prev}_i^{(t)}, i \in [d]$ are the previous rebuild times (Definition \ref{defn6.12}) for the branching tree train.

The algorithm succeeds w.h.p with total runtime $(m + Q) m^{o(1)}$ for $Q ~ \stackrel{\text{def}}{=} \sum_{t = 1}^\tau |U^{(t)}| \leq \text{poly}(n).$ Also, levels $i, i + 1, \ldots, d$ of the branching tree train can be rebuilt at any point in $m^{1 + o(1)}/k^i$ time. 
    
\end{thm}

\begin{thm}\label{thm6.2} (\cite{main}) There is a data structure that on a dynamic graph $G^{(t)}$ maintains a collection of $s = O(\log n)^d$ spanning trees $T_1, T_2, \ldots, T_s \subseteq G^{(t)}$ for $d = O(\log^{1/8} m)$, and supports the following operations:

\begin{itemize}
    \item UPDATE($U^{(t)}, \mathbf{g}^{(t)}, \mathbf{\ell}^{(t)}$): Update the gradients and lengths to $\mathbf{g^{(t)}}$ and $\mathbf{\ell}^{(t)}$. For the update to be supported, we require that $U^{(t)}$ contains only edge insertions/deletions and $\mathbf{g}^{(t)}, \mathbf{\ell}^{(t)}$ and $U^{(t)}$ satisfy the hidden stable-flow chasing property (Definition \ref{defn6.1}) with hidden circulation $\mathbf{c}^{(t)}$ and upper bounds $\mathbf{w}^{(t)}$, and for a parameter $\alpha$, 

    $$\frac{\langle \mathbf{g}^{(t)}, \mathbf{c}^{(t)} \rangle}{||\mathbf{w}^{(t)}||_1} \leq \alpha.$$

    \item QUERY(): Returns a tree $T_i$ for $i \in [s]$ and a cycle $\Delta$ represented as $m^{o(1)}$ paths on $T_i,$ (specified by their endpoints and the tree index) and $m^{o(1)}$ explicitly given off-tree edges such that for $\kappa = \exp(-O(\log^{7/8} m \cdot \log \log m)),$

    $$\frac{\langle \mathbf{g}^{(t)}, \Delta \rangle}{||\mathbf{L}^{(t)} \Delta||_1} \leq -\kappa \alpha.$$

    Over $\tau$ stages the algorithm succeeds w.h.p with total run time $m^{o(1)} (m + Q)$ for $Q = \sum_{t \in [\tau]} |U^{(t)}|.$
\end{itemize}
    
\end{thm}

\begin{notn} Following the notation of \cite{main}, we denote the data structure described in Theorem \ref{thm6.2} as $\mathcal{D}^{(HSFC)}.$
    
\end{notn}

\subsection{Rebuilding Game}\label{rebuilding_subsection}

Now, we analyze the time complexity of the Rebuilding Game, which we know is given as $O\left(\frac{C_r Kd}{\gamma_g}(m + T)\right)$ in Lemma \ref{lem8.3}. At the end of Section 8.2 of \cite{main}, when describing how to use the Rebuilding game to get the main data structure described in Theorem \ref{thm6.2}, the authors state that $d = \sqrt[8]{\log m}, k = \exp(O(\log^{7/8} m)), \gamma_g = \exp(-O(\log^{7/8} m \log \log m)), C_r = \exp(O(\log^{7/8} m \log \log m)), T = (m + Q) \exp(O(\log^{7/8} m \log \log m)).$ Plugging these values in, we get the bound $(m + Q) \exp(O(\log^{7/8} m \log \log m))$ given in \cite{main}.

Here, $Q = \sum_{t \in \tau} |U^{(t)}|$ as defined in Theorem \ref{thm6.2}, and we need to bound $Q$ in terms of $m$. To do this, we turn to Lemma 9.4 from \cite{main}, which we reproduce below, where Algorithm 7 is the name \cite{main} gives to their overall algorithm:

\begin{lem}\label{lem9.4} (\cite{main}) Consider a call to MINCOSTFLOW (Algorithm 7) and let $U^{(t)}$ be as in line 20 [of Algorithm 7]. Then $\sum_t |U^{(t)}| \leq \tilde O(m \kappa^{-2} \alpha^{-2} \varepsilon^{-1}).$
    
\end{lem}

Thus, based on Lemma \ref{lem9.4}, we have $Q = \tilde O(m \kappa^{-2} \alpha^{-2} \varepsilon^{-1}).$ Here, $\kappa = \exp(-O(\log^{7/8} m \log \log m))$ as defined in Theorem \ref{thm6.2} and $\alpha = \frac{1}{1000 \log mU}$ where $U$ is as defined in Definiton \ref{vars_defn}. Further, $\varepsilon$ is as defined in the DETECT operation of Lemma \ref{dynamic_trees_lemma}. We treat $U$ and $\varepsilon$ as constants that we absorb into the $O$ notation.

In order to dissect what is being included in the $\tilde O$ notation above for $Q$, we refer to Section 4 of \cite{main}, specifically Theorem \ref{thm4.3} and the text immediately following it, where the authors in \cite{main} state that a certain operation decreases the potential $\Phi(f)$ (see Definition \ref{defn_potential}) by $\Omega(\kappa^2)$. The potential drops from $O(m \log mU)$ to $-O(m \log m),$ so this takes $O(m \log m \kappa^{-2})$ operations, where $\kappa^{-2} = \exp(O(\log^{7/8} m \log \log m))$. We also note that $\alpha^{-2} = O(\log^2 m).$

Now we do some big $O$ computation. Letting $c_i$ denote constants, we have:

$$Q = O(m \log m \kappa^{-2} \alpha^{-2}) \leq c_1 m \log^3 m \exp(O(\log^{7/8} m \log \log m))$$ $$\leq \exp\left(c_2 \log^{7/8} m \log \log m + \log m + 3 \log \log m + \log c_1 \right) = m \exp(O(\log^{7/8} m \log \log m)).$$

The last step follows by pulling out the $e^{\log m} = m$ term and then noticing that $\log^{7/8} m \log \log m$ dominates $3 \log \log m$ and $\log c_1.$

From this, we get:

$$(m + Q)\exp(O(\log^{7/8} m \log \log m)) = m \exp(O(\log^{7/8} m \log \log m))$$ $$+ m \exp(O(\log^{7/8} m \log \log m)) \exp(O(\log^{7/8} m \log \log m)) = m \exp(O(\log^{7/8} m \log \log m)).$$

As a sanity check, we note that $\exp(O(\log^{7/8} m \log \log m)) = m^{o(1)}$ because:

$$\lim_{m \to \infty}\frac{c_1 \log^{7/8} m \log \log m}{ \log m} = \lim_{m\to \infty} \frac{c_1 \log \log m}{\log^{1/8} m} = 0.$$

Now we have the following mathematical lemma:

\begin{lem}\label{lem_limit} Suppose $f(m)$ and $g(m)$ are positive functions that both go to $\infty$ as $m \to \infty$ such that $f(m) = o(g(m)).$ Then $e^{f(m)} = o(e^{g(m)})$.
    
\end{lem}

The proof of Lemma \ref{lem_limit} is rather obvious: $\forall \varepsilon > 0, ~ \exists m$ such that $f(m) \leq \varepsilon \cdot g(m).$ Thus $e^{f(m) - g(m)} \leq e^{(\varepsilon - 1)g(m)}.$ Then, since $\varepsilon < 1,$ we clearly have $\lim_{m \to \infty} e^{(\varepsilon - 1)g(m)} = 0.$

Applying this logic above, it follows that $\lim_{m \to \infty} \frac{\exp(O(\log^{7/8} m \log \log m))}{m^{\frac{1}{\log \log m}}} = 0$ so we indeed have that $\exp(O(\log^{7/8} m \log \log m)) = m^{o(1)}.$

\subsection{MinCostFlow Algorithm}

In this subsection, we precisely calculate the time complexity for the MinCostFlow Algorithm. This, combined with the previous subsection, allows us to establish the overall time complexity in the next subsection.

In Section 9 of \cite{main}, the authors show that the MINCOSTFLOW Algorithm satisfies the hypotheses of Theorem \ref{thm4.3}. Thus, we conclude that the MINCOSTFLOW Algorithm runs for $\tilde O(m \kappa^{-2} \alpha^{-2})$ iterations, and following the same logic as in Section \ref{rebuilding_subsection}, we can rewrite this as $m \exp(O(\log^{7/8} m \log \log m)).$

For the time per iteration, we note that there are $s = O((\log n)^d) = O\left((\log m)^{\sqrt[8]{\log m}}\right)$ trees maintained by the $\mathcal{D}^{(HSFC)}$ data structure. For each of these trees, we invoke the Dynamic Tree data structure operations, which take $O(\log m)$ based on the guarantees of Lemma \ref{dynamic_trees_lemma}. 

Therefore, we get a bound of $m \exp(O(\log^{7/8} m \log \log m)) \cdot O(\log m) \cdot O\left((\log m)^{\sqrt[8]{\log m}}\right)$. We note that $(\log m)^{1 + \sqrt[8]{\log m}} = \exp((1 + \sqrt[8]{\log m}) \log \log m) = \exp(O(\log^{7/8} m \log \log m)).$ Thus, we still have $m \exp(O(\log^{7/8} m \log \log m))$ as our time complexity.






\subsection{Comparison with Edmonds Karp}

The Edmonds Karp algorithm is another well known algorithm for solving max flow, which runs in $O(mn^2) = O(m^3)$ time. We now compare the growth of $m^3$ vs $m \exp(C \log^{7/8} m \log \log m)$ for large $m$ and constant $C$. The goal is to find out when $m^3$ becomes larger than $m \exp(C \log^{7/8} m \log \log m)$.

We have:

$$m^3 > m \exp(\log^{7/8} m \log \log m) \implies 2 \log^{1/8} m > C \log \log m.$$

Letting $x = \log m,$ we need to solve $2 \sqrt[8]{x} > C \log x.$ If we take $C = 1,$ Wolfram Alpha indicates this inequality holds for all $x > 3.03 \cdot 10^7$, which means $m > \exp(3.03 \cdot 10^7) = 1.01 \cdot 10^{13154921}$. If the number of edges/vertices of the graph were to exceed this number, then Edmonds Karp would be faster. If $C = 2,$ the same calculations give us $m > 3.32 \cdot 10^{93334255463}.$

It is also worth considering how much slower this algorithm is for a more reasonable value of $m$. Note that because we do not know the value of the constant $C$, it is unreasonable to analyze small values of $m$ (like $10$ or $100$). Taking $C = 1,$ if we assume our graph has $10^{10}$ edges, then we have:

$$\frac{10^{10} \exp(\log^{7/8}(10^{10}) \log \log 10^{10})}{10^{30}} \approx 15.58.$$

Similar to the behavior of the function in Section \ref{ipm_implementation}, this function gets larger before it gets smaller. When $m = 10^{1000},$ it equals $2.95 \cdot 10^{941},$ and of course these numbers would be larger for a larger $C.$ Thus we can conclude that for any reasonably sized graph (where reasonably sized is defined as can fit in the storage of the world's largest supercomputers), a traditional algorithm like Edmonds-Karp will be faster.

\section{Conclusion}\label{conclusion}

\subsection{Summary}

We began with preliminaries and provided a detailed list of all of the different subparts of the algorithm in \cite{main} that would have to be implemented in order for the main algorithm to be implemented. After this, we discussed our implementations of portions of the algorithm, including the Low Stretch Decomposition, the Rebuilding Game and the Interior Point Method with insights into why certain implementation decisions were made. Finally, we examined the overall time complexity of the algorithm in detail and worked out a more precise asymptotic bound. 

\subsection{Future Directions of Study}

Several portions of the algorithm in \cite{main} remain unimplemented, including primarily the expander decomposition and a fully correct Interior Point Method. The nested list provided in Section \ref{layout} combined with the stubs provided in the GitHub should give a roadmap for any future attempts at implementing this algorithm. Separately, while the algorithm in \cite{main} is the asymptotically fastest algorithm for max flow to date, as discussed in the final subsection it is not designed to be practical. Another interesting area of research would be to attempt to find the fastest algorithm which can be implemented reasonably or prove that existing algorithms already meet those criteria.
 
\bibliography{Index}
\end{document}